\documentclass[final,5p,times,twocolumn]{elsarticle}
\usepackage{lineno,hyperref,amsmath,amsthm,amssymb,amsfonts,ragged2e,color,subfig}
\journal{``Plasma Physics Reports''}
\begin{document}
\begin{frontmatter}
\title{Dust-ion-acoustic shock waves in magnetized plasma having super-thermal electrons}
\author{T. Yeashna$^{1,*}$, R.K. Shikha$^{1,**}$, N.A. Chowdhury$^{2,***}$, A. Mannan$^{1,\dag}$, S. Sultana$^{1,\ddag}$, A.A. Mamun$^{1,\S}$}
\address{$^1$ Department of Physics, Jahangirnagar University, Savar, Dhaka-1342, Bangladesh\\
$^2$ Plasma Physics Division, Atomic Energy Centre, Dhaka-1000, Bangladesh\\
e-mail: $^*$yeashna147phy@gmail.com $^{**}$shikha261phy@gmail.com, $^{***}$nurealam1743phy@gmail.com,\\
$^{\dag}$abdulmannan@juniv.edu, $^{\ddag}$ssultana@juniv.edu, $^{\S}$mamun\_phys@juniv.edu}
\begin{abstract}
The propagation of dust-ion-acoustic shock waves (DIASHWs) in a three-component magnetized
plasma having inertialess super-thermal electrons, inertial warm positive
ions and negative dust grains has been investigated. A Burgers' equation is derived by employing the reductive perturbation method.
Under consideration of inertial warm positive ions and negative dust grains, both positive and negative shock structures
are numerically observed in the presence of super-thermal electrons.
The effects of oblique angle ($\delta$), spectral index ($\kappa$), kinematic viscosity ($\eta$),  number density and charge state of the
plasma species on the formation of the DIASHWs are examined. It is found that the positive and
negative shock wave potentials increase with the oblique angle.
It is also observed that the magnitude of the amplitude of positive and negative shock waves is not affected by the variation
of the kinematic viscosity of plasma species but the steepness of the positive and negative shock waves decreases with kinematic viscosity of plasma species. The implications of our findings in space and laboratory plasmas are briefly discussed.
\end{abstract}
\end{frontmatter}
\section{Introduction}
\label{2sec:introduction}
Dust grains are omnipresent components in most of the astrophysical environments, viz., interstellar clouds \cite{Spitzer1978,Savage1979,Herbst1995,C1}, circumstellar clouds \cite{Ossenkopf1992,Andrews2005,Dijkstra2005}, interplanetary space \cite{Mackinnon1987,Sandford1989,Liou1996},
Earth's magnetosphere \cite{Horanyi1987}, Saturn's B ring \cite{Northrop1983} and magnetosphere \cite{Humes1980,Northrop1983a},
Jupiter's magnetosphere \cite{Northrop1983a,Humes1974}, cometary tails \cite{Mazets1987,Gurnett1986}, and also in laboratory plasmas \cite{Selwyn1989,Boufendi2002}. In a realistic dusty plasma medium (DPM), dust grains can collide frequently with lighter electrons than the
ions, and due to these frequent collisions, dust grains are generally negatively charged \cite{Northrop1982,Merlino1998} massive object by collecting
electrons from the surrounding environments \cite{Shukla1992}. It was observed that dust grains are responsible to change the dynamics
of the DPM \cite{Ghosh2000,Nakamura2002,Mamun2010,Samsonov2004,Bacha2012}.

It is now well established that the Earth's magnetosphere plasma sheet \cite{Vasyliunas1968,Lui1983,Williams1988,Christon1988},
the solar wind \cite{Abraham-Shrauner1977,Gosling1981}, Jupiter's magnetosphere \cite{Divine1983},
and Saturn's magnetosphere \cite{Armstrong1983} plasma contain highly energetic particles which can exhibit
high energy tails. Maxwellian distribution is only appropriate for describing iso-thermal particles
but it fails to describe the dynamics of these highly energetic particles. The super-thermal kappa ($\kappa$)
distribution was first introduced by Vasyliunas \cite{Vasyliunas1968} for congruent
description of these energetic particles. The parameter $\kappa$ in the $\kappa$ distribution
is called the spectral index which describes the deviation of the plasma particles from the thermally
equilibrium state. The $\kappa$-distribution reduces to the ordinary Maxwellian distribution for large values of $\kappa$ (i.e., $\kappa\rightarrow\infty$).
Shah \textit{et al.} \cite{Shah2012} investigated the ion-acoustic (IA) shock waves (IASHWs) in the presence of super-thermal electrons, and observed that the height of the positive shock profile increases with an increase in the value of spectral index
of electrons. Adnan \textit{et al.} \cite{Adnan2014} studied small amplitude IA solitons in a magnetized
plasma having $\kappa$-distributed electrons, and found that the amplitude of the positive solitons increases
with the electrons super-thermality. Haider \textit{et al.} \cite{Haider2017} analysed dust-ion-acoustic (DIA) solitary waves (DIASWs) and DIA shock
waves (DIASHWs) in the presence of super-thermal electrons, and demonstrated that positive potential associated with DIASWs
increases with the increase of super-thermality of the electrons.

Shock wave mostly arises in nature due to balance between nonlinear and dissipative
forces \cite{Zel'dovich1967}. The collision between charged and neutral particles, kinematic
viscosity of the medium components, and Landau damping are to be responsible for dissipation \cite{Sahu2014}.
The presence of kinematic viscosity plays a major role in generating nonlinear waves \cite{Hossen2017}.
Bansal \textit{et al.} \cite{Bansal2020} analysed IASHWs in DPM having super-thermal electrons,
and observed that the steepness of shock structures decreases with the increase of kinematic
viscosity of plasma species. Borah \textit{et al.} \cite{Borah2016} studied DIASWs and DIASHWs in a three-component DPM,
and reported that the steepness of the shock structures reduces with the increase of the kinematic viscosity of the plasma component.

A number of authors considered an external magnetic field to investigate nonlinear electrostatic
shock \cite{Hossen2017,Kamran2021} and solitary \cite{El-Monier2019} waves in different plasma medium.
Hossen \textit{et al.} \cite{Hossen2017} studied the electrostatic shock structures in magnetized DPM,
and found that the magnitude of the positive and negative shock profiles increases with the oblique
angle ($\delta$) which arises due to the external magnetic field. El-Monier \textit{et al.} \cite{El-Monier2019}
investigated nonlinear IA solitary structures in a three-component magnetized plasma and highlighted that the
positive potential increases with the increase of oblique angle. Kamran \textit{et al.} \cite{Kamran2021}
analysed DIASHWs in a three-component plasma composed of $\kappa$-distributed electrons, mobile ions,
and stationary dust grains, and observed that the magnitude of the amplitude of negative shocks increases
with the increase of electron number density. To the best knowledge of the authors, no attempt has been made
to study the DIASHWs in a three-component magnetized DPM by considering kinematic viscosities
of both inertial warm positive ions and negative dust grains, and inertialess super-thermal electrons. The aim of the present
investigation is, therefore, to derive Burgers' equation and investigate DIASHWs in a three-component magnetized DPM,
and to observe the effects of various plasma parameters on the configuration of DIASHWs.

This manuscript is organized as follows: The governing equations are
described in section \ref{2sec:Governing Equations}. The Burgers' equation is derived
in section \ref{2sec:Derivation of the Burgers' equation}. Numerical analysis is reported
in section \ref{2sec:Numerical analysis}. A brief
conclusion of our present work is provided in section \ref{2sec:Conclusion}.
\section{Governing Equations}
\label{2sec:Governing Equations}
We present a simplified fluid model for DIASHWs
in a magnetized, three-component DPM consisting
of inertial negatively charged dust particles, positively
charged warm ions, and inertialess $\kappa$-distributed electrons.
An external magnetic field $\mathbf{B}_0$ has been considered in the system directed along
the $z$-axis signifying $\mathbf{B}_0 = B_0\hat{z}$ where $B_0$ and $\hat{z}$ represent the
strength of the external magnetic field and unit vector directed along the $z$-axis, respectively.
The wave propagation vector is assumed to produce an oblique angle ($\delta$) with the external magnetic field.
The dynamics of the magnetized DPM is governed by the following equations:
\begin{eqnarray}
&&\hspace*{-1.3cm}\frac{\partial \tilde{n}_{d}}{\partial \tilde{t}}+\acute{\nabla}\cdot(\tilde{n}_d \tilde{u}_d)=0,
\label{2eq:1}\\
&&\hspace*{-1.3cm}\frac{\partial \tilde{u}_d}{\partial\tilde{t}}+(\tilde{u}_d\cdot\acute{\nabla})\tilde{u}_{d}=\frac{Z_de}{m_d}\acute{\nabla}\tilde{\psi} -\frac{Z_deB_0}{m_d}(\tilde{u}_{d}\times\hat{z})+\tilde{\eta}_d\acute{\nabla}^2\tilde{u}_d,
\label{2eq:2}\\
&&\hspace*{-1.3cm}\frac{\partial\tilde{n}_i}{\partial \tilde{t}}+\acute{\nabla}\cdot(\tilde{n}_i\tilde{u}_i)=0,
\label{2eq:3}\\
&&\hspace*{-1.3cm}\frac{\partial\tilde{u}_i}{\partial \tilde{t}}+(\tilde{u}_i\cdot\acute{\nabla})\tilde{u}_{i}=-\frac{Z_ie}{m_i}\acute{\nabla}\tilde{\psi} +\frac{Z_ieB_0}{m_i}(\tilde{u}_i\times\hat{z})
\nonumber\\
&&\hspace*{3cm}-\frac{1}{m_i\tilde{n}_i}\acute{\nabla} P_i+\tilde{\eta}_i\acute{\nabla}^2\tilde{u}_i,
\label{2eq:4}\\
&&\hspace*{-1.3cm}\acute{\nabla}^2\tilde{\psi}=4\pi e[\tilde{n}_e-Z_i\tilde{n}_i+Z_d\tilde{n}_d],
\label{2eq:5}\
\end{eqnarray}
where $\tilde{n}_d$ ($\tilde{n}_i$) is the dust (ion) number density, $m_d$ ($m_i$) is
the dust (ion) mass, $Z_d$ ($Z_i$) is the charge state of the dust (ion),
$e$ is the magnitude of electron charge, $\tilde{u}_d$ ($\tilde{u}_i$) is the dust (ion)
fluid velocity, $\tilde{\eta}_d = \mu_d/m_d\tilde{n}_d$ ($\tilde{\eta}_i = \mu_i/m_i\tilde{n}_i$) is the kinematic viscosity of the dust (ion), $P_i$ is the pressure of positive ion, and $\tilde{\psi}$ represents the electrostatic wave potential.
Now, we  are introducing normalized parameters, namely, $n_d\rightarrow\tilde{n}_d/n_{d0}$, $n_i\rightarrow\tilde{n}_i/n_{i0}$, and $n_e\rightarrow\tilde{n}_e/n_{e0}$,
where  $n_{d0}$, $n_{i0}$, and $n_{e0}$ are the equilibrium number densities of the negative dust grains, positive ions, and electrons, respectively;
$u_d\rightarrow\tilde{u}_d/C_i$, $u_i\rightarrow\tilde{u}_i/C_i$ [where $C_i=(Z_ik_BT_e/m_i)^{1/2}$, $k_B$ being the Boltzmann constant, and $T_e$ being temperature of the electron]; $\psi\rightarrow\tilde{\psi}e/k_BT_e$; $t=\tilde{t}/\omega_{p}^{-1}$ [where $\omega_{p}^{-1}=(m_i/4\pi e^{2}Z_i^{2}n_{i0})^{1/2}$]; $\nabla=\acute{\nabla}/\lambda_{D}$ [where $\lambda_{D}=(k_BT_e/4\pi e^2Z_in_{i0})^{1/2}$]. The pressure term of the positive ions can be recognized as $P_i=P_{i0}(\tilde{n}_i/n_{i0})^\gamma$ with $P_{i0}=n_{i0}k_BT_i$ being the equilibrium
pressure of the positive ions, and $T_i$ is the temperature of warm positive ions, and
$\gamma=(N+2)/N$ (where $N$ is the degree of freedom and for three-dimensional case
$N=3$, then $\gamma=5/3$). For simplicity, we have considered ($\tilde{\eta}_d\approx\tilde{\eta}_i=\eta$), and $\eta$ is
normalized by $\omega_{p}\lambda_D^{2}$. The quasi-neutrality condition at equilibrium for our plasma model
can be written as $n_{e0} + Z_dn_{d0} \approx Z_in_{i0}$. Equations \eqref{2eq:1}$-$\eqref{2eq:5} can be
expressed in the normalized form as:
\begin{eqnarray}
&&\hspace*{-1.3cm}\frac{\partial n_{d}}{\partial t}+\nabla\cdot(n_{d} u_{d})=0,
\label{2eq:6}\\
&&\hspace*{-1.3cm}\frac{\partial u_{d}}{\partial t}+(u_{d}\cdot\nabla)u_{d}=\alpha_1\nabla\psi -\alpha_1\Omega_{c}(u_{d}\times\hat{z})+\eta\nabla^2u_d,
\label{2eq:7}\\
&&\hspace*{-1.3cm}\frac{\partial n_{i}}{\partial t}+\nabla \cdot(n_{i} u_{i})=0,
\label{2eq:8}\\
&&\hspace*{-1.3cm}\frac{\partial u_{i}}{\partial t}+(u_{i}\cdot\nabla)u_{i}=-\nabla\psi +\Omega_{c}(u_{i}\times\hat{z})-\alpha_2\nabla n_{i}^{\gamma-1}+\eta\nabla^2u_i,
\label{2eq:9}\\
&&\hspace*{-1.3cm}\nabla^2 \psi=n_d(1-\mu_e)-n_i+n_e\mu_e,
\label{2eq:10}\
\end{eqnarray}
other plasma parameters are defined as $\alpha_1=Z_dm_i/Z_im_d$,
$\alpha_2=\gamma T_i/(\gamma-1)Z_iT_e$, $\mu_e = n_{e0}/Z_in_{i0}$,
and $\Omega_c=\omega_{ci}/\omega_{P}$ [where $\omega_{ci}=Z_ieB_0/m_i$].
Now, the expression for the number density of electrons following the
$\kappa$-distribution can be written as \cite{Haider2017,Adnan2014}
\begin{eqnarray}
&&\hspace*{-1.3cm}n_e=\left[1 -\frac{\psi}{\kappa-3/2}\right]^{-\frac{\kappa+1}{2}},
\label{2eq:11}\
\end{eqnarray}
where the parameter $\kappa$ represents the non-thermal properties of the electrons.
Now, by expanding  Eq. \eqref{2eq:11} up to third order in $\psi$, and substituting in Eq. \eqref{2eq:10}, we can write
\begin{eqnarray}
&&\hspace*{-1.3cm}\nabla^2 \psi=\mu_e+n_d(1-\mu_e)-n_i+\sigma_1\psi
\nonumber\\
&&\hspace*{-0.2cm}+\sigma_2\psi^2+\sigma_3\psi^3+\cdot\cdot\cdot,
\label{2eq:12}\
\end{eqnarray}
where
\begin{eqnarray}
&&\hspace*{-1.3cm}\sigma_1=\mu_e[(\kappa+1)/2(\kappa-3/2)],
\nonumber\\
&&\hspace*{-1.3cm}\sigma_2=\mu_e[(\kappa+1)(\kappa+3)/8(\kappa-3/2)^2],
\nonumber\\
&&\hspace*{-1.3cm}\sigma_3=\mu_e[(\kappa+1)(\kappa+3)(\kappa+5)/48(\kappa-3/2)^3].
\nonumber\
\end{eqnarray}
We note that the terms containing $\sigma_1$, $\sigma_2$, and $\sigma_3$ are
the contribution of $\kappa$-distributed electrons.
\section{Derivation of the Burgers' equation}
\label{2sec:Derivation of the Burgers' equation}
To derive the Burgers' equation for the DIASHWs propagating in a magnetized plasma, we are going to employ reductive perturbation
method \cite{C2}. First we introduce the stretched co-ordinates \cite{Hossen2017,Washimi1966}
\begin{eqnarray}
&&\hspace*{-1.3cm}\xi=\epsilon(l_xx+l_yy+l_zz-v_p t),
\label{2eq:13}\\
&&\hspace*{-1.3cm}\tau={\epsilon}^2 t,
\label{2eq:14}\
\end{eqnarray}
where $v_p$ express the phase speed and $\epsilon$ represent a small parameter ($0<\epsilon<1$).
The $l_x$, $l_y$, and $l_z$ (i.e., $l_x^2+l_y^2+l_z^2=1$) are the directional cosines of the wave
vector $k$ along $x$, $y$, and $z$-axes, respectively. Then, the dependent variables can be expressed
in power series of $\epsilon$ as
\begin{eqnarray}
&&\hspace*{-1.3cm}n_{d}=1+\epsilon n_{d}^{(1)}+\epsilon^2 n_{d}^{(2)}+\epsilon^3 n_{d}^{(3)}+\cdot\cdot\cdot,
\label{2eq:15}\\
&&\hspace*{-1.3cm}n_{i}=1+\epsilon n_{i}^{(1)}+\epsilon^2 n_{i}^{(2)}+\epsilon^3 n_{i}^{(3)}+\cdot\cdot\cdot,
\label{2eq:16}\\
&&\hspace*{-1.3cm}u_{dx,y}=\epsilon^2 u_{dx,y}^{(1)}+\epsilon^3 u_{dx,y}^{(2)}+\cdot\cdot\cdot,
\label{2eq:17}\\
&&\hspace*{-1.3cm}u_{ix,y}=\epsilon^2 u_{ix,y}^{(1)}+\epsilon^3 u_{ix,y}^{(2)}+\cdot\cdot\cdot,
\label{2eq:18}\\
&&\hspace*{-1.3cm}u_{dz}=\epsilon u_{dz}^{(1)}+\epsilon^2 u_{dz}^{(2)}+\cdot\cdot\cdot,
\label{2eq:19}\\
&&\hspace*{-1.3cm}u_{iz}=\epsilon u_{iz}^{(1)}+\epsilon^2 u_{iz}^{(2)}+\cdot\cdot\cdot,
\label{2eq:20}\\
&&\hspace*{-1.3cm}\psi=\epsilon \psi^{(1)}+\epsilon^2\psi^{(2)}+\cdot\cdot\cdot.
\label{2eq:21}\
\end{eqnarray}
Now, by substituting Eqs. \eqref{2eq:13}$-$\eqref{2eq:21} in Eqs. \eqref{2eq:6}$-$\eqref{2eq:9} and \eqref{2eq:12},
we obtain a set of first-order equations in the following form
\begin{eqnarray}
&&\hspace*{-1.3cm}n_d^{(1)} = -\frac{ \alpha_1l_z^2}{v_p^2}\psi^{(1)},
\label{2eq:22}\\
&&\hspace*{-1.3cm}u_{dz}^{(1)} = -\frac{\alpha_1l_z}{v_p}\psi^{(1)},
\label{2eq:23}\\
&&\hspace*{-1.3cm}n_i^{(1)} = \frac{3l_z^2}{\big(3v_p^2 - 2\alpha_2l_z^2 \big)}\psi^{(1)},
\label{2eq:24}\\
&&\hspace*{-1.3cm}u_{iz}^{(1)} = \frac{3v_p l_z}{\big(3v_p^2 - 2\alpha_2l_z^2\big)}\psi^{(1)}.
\label{2eq:25}\
\end{eqnarray}
Now, the phase speed of DIASHWs can be written as
\begin{eqnarray}
&&\hspace*{-1.3cm}v_{p}\equiv v_{p+} = l_z\sqrt{{\frac{-a_1+\sqrt{a_1^2-12\sigma_1a_2}}{6\sigma_1}}},
\label{2eq:26}\\
&&\hspace*{-1.3cm}v_{p}\equiv v_{p-} =l_z\sqrt{{\frac{-a_1-\sqrt{a_1^2-12\sigma_1a_2}}{6\sigma_1}}},
\label{2eq:27}\
\end{eqnarray}
where $a_1=3\mu_e\alpha_1-2\alpha_2\sigma_1-3-3\alpha_1$ and $a_2=2\alpha_1\alpha_2+2\mu_e\alpha_1\alpha_2$.
The $x$ and $y$-components of the first-order momentum equations can be manifested as
\begin{eqnarray}
&&\hspace*{-1.3cm}u_{dx}^{(1)}=-\frac{l_y}{\Omega_{c}} \frac{\partial\psi^{(1)}}{\partial\xi},
\label{2eq:28}\\
&&\hspace*{-1.3cm}u_{dy}^{(1)} =  \frac{l_x}{\Omega_{c}} \frac{\partial\psi^{(1)}}{\partial\xi},
\label{2eq:29}\\
&&\hspace*{-1.3cm}u_{ix}^{(1)}=-\frac{3l_y v_p^2}{\Omega_{c}(3v_p^2 - 2\alpha_2l_z^2)} \frac{\partial\psi^{(1)}}{\partial\xi},
\label{2eq:30}\\
&&\hspace*{-1.3cm}u_{iy}^{(1)} =  \frac{3l_x v_p^2}{\Omega_{c}(3v_p^2 - 2\alpha_2l_z^2)} \frac{\partial\psi^{(1)}}{\partial\xi}.
\label{2eq:31}\
\end{eqnarray}
\begin{figure}
\centering
\includegraphics[width=80mm]{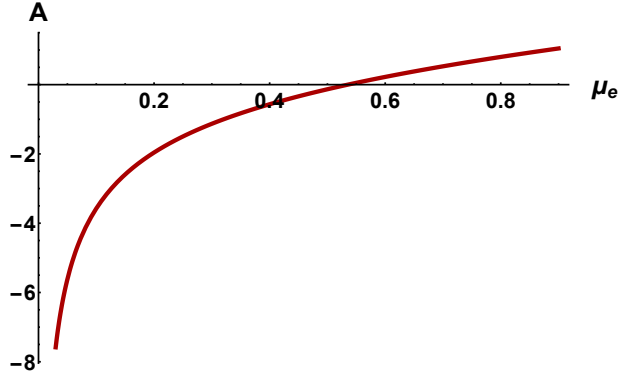}
\caption{The variation of nonlinear coefficient $A$ with $\mu_e$. Other plasma parameters are
$\alpha_1 = 10^{-3}$, $\alpha_2 =0.03$, $\delta = 10^\circ$, $\kappa = 2$, and $v_p\equiv v _{p+}$.}
 \label{2Fig:F1}
\end{figure}
Now, by taking the next higher-order terms, the equation of
continuity, momentum equation, and Poisson’s equation can be written as
\begin{eqnarray}
&&\hspace*{-1.3cm} \frac{ \partial n_d^{(1)}}{\partial \tau} - v_p  \frac{ \partial n_d^{(2)}}{\partial \xi}
+ l_x \frac{ \partial u_{dx}^{(1)}}{\partial \xi} + l_y \frac{ \partial u_{dy}^{(1)}}{\partial \xi}+ l_z \frac{ \partial u_{dz}^{(2)}}{\partial \xi}
\nonumber\\
&&\hspace*{1.8cm}+ l_z \frac{\partial}{\partial \xi} \Big(n_d^{(1)}  u_{dz}^{(1)} \Big) = 0,
\label{2eq:32}\\
&&\hspace*{-1.3cm} \frac{ \partial u_{dz}^{(1)}}{\partial \tau} - v_p \frac{ \partial u_{dz}^{(2)}}{\partial \xi} + l_z  u_{dz}^{(1)} \frac{\partial u_{dz}^{(1)}}{\partial \xi} - \alpha_1 l_z \frac{\partial \psi^{(2)}}{\partial \xi}- \eta \frac{\partial^2 u_{hz}^{(1)}}{\partial \xi^2} = 0,
\label{2eq:33}\\
&&\hspace*{-1.3cm} \frac{ \partial n_i^{(1)}}{\partial \tau} - v_p  \frac{ \partial n_i^{(2)}}{\partial \xi}
+ l_x \frac{ \partial u_{ix}^{(1)}}{\partial \xi} + l_y \frac{ \partial u_{iy}^{(1)}}{\partial \xi} + l_z \frac{ \partial u_{iz}^{(2)}}{\partial \xi}
\nonumber\\
&&\hspace*{1.8cm}+ l_z \frac{\partial}{\partial \xi} \Big(n_i^{(1)}  u_{iz}^{(1)} \Big) = 0,
\label{2eq:34}\\
&&\hspace*{-1.3cm} \frac{ \partial u_{iz}^{(1)}}{\partial \tau} - v_p \frac{ \partial u_{iz}^{(2)}}{\partial \xi} +  l_z u_{iz}^{(1)}\frac{\partial u_{iz}^{(1)}}{\partial \xi} +l_z \frac{\partial \psi^{(2)}}{\partial \xi}
\nonumber\\
&&\hspace*{0.5cm}+ \alpha_2l_z \Biggr[ \frac{2}{3}\frac{\partial n_i^{(2)}}{\partial \xi}
-\frac{1}{9}  \frac{\partial n_{i}^{(1)^2}}{\partial \xi} \Biggr] - \eta \frac{\partial^2 u_{iz}^{(1)}}{\partial \xi^2} = 0,
\label{2eq:35}\\
&&\hspace*{-1.3cm}n_i^{(2)}=(1-\mu_e)n_d^{(2)}+\sigma_1\psi^{(2)}+\sigma_2\psi^{(1)^2}.
\label{2eq:36}\
\end{eqnarray}
Finally, the next higher-order terms of Eqs. \eqref{2eq:6}-\eqref{2eq:9} and \eqref{2eq:12},
with the help of Eqs. \eqref{2eq:22}-\eqref{2eq:36}, can provide the Burgers' equation
\begin{eqnarray}
&&\hspace*{-1.3cm} \frac{\partial\Psi}{\partial\tau} + A\Psi\frac{\partial\Psi}{\partial\xi}=C\frac{\partial^2\Psi}{\partial\xi^2},
\label{2eq:37}\
\end{eqnarray}
where $\Psi=\psi^{(1)}$ is used for simplicity. The nonlinear coefficient ($A$) and dissipative coefficient ($C$) are represented, respectively, as
\begin{eqnarray}
&&\hspace*{-1.3cm}A=\frac{2\sigma_2s_1^3v_p^4+3\alpha_1^2s_1^3l_z^4(1-\mu_e)-81v_p^6l_z^4-F_1}{2\alpha_2v_pl_z^2s_1^3(\mu_e-1)+F_2},
\label{2eq:38}\\
&&\hspace*{-1.3cm}C=\frac{\eta}{2},
\label{2eq:39}\
\end{eqnarray}
where $F_1 =6\alpha_2s_1v_p^4l_z^6$, $F_2 =-18s_1l_z^2v_p^5$, and $s_1 = 3v_p^2 - 2\alpha_2l_z^2$.
To obtain stationary shock wave solution of this Burgers' equation, we consider a frame of reference
that advances with shock speed $U_0$. The space ($\zeta$) and time ($\tau$) coordinates in such frame
are expressed as $\zeta =\xi-U_0\tau'$ and $\tau =\tau'$. These allow us
to write the stationary shock wave solution as \cite{Karpman1975,Hasegawa1975}
\begin{eqnarray}
&&\hspace*{-1.3cm}\Psi=\Psi_m\Big[1 - \tanh\bigg(\frac{\zeta}{\Delta}\bigg)\Big],
\label{2eq:40}\
\end{eqnarray}
where the amplitude $\Psi_m$ and the width $\Delta$ is given by
\begin{eqnarray}
&&\hspace*{-1.3cm}\Psi_m=\frac{U_0}{A},~~\text{and}~~\Delta=\frac{2C}{U_0}.
\label{2eq:41}\
\end{eqnarray}
It can be seen from Eq. \eqref{2eq:33} that the amplitude of the shock wave becomes infinity
corresponding to the value of $A = 0$ (due to $C>0$ and $U_0>0$), and this refers to that our theory
(specially, the RPM) is only valid for the small amplitude waves.
\begin{figure}
\centering
\includegraphics[width=80mm]{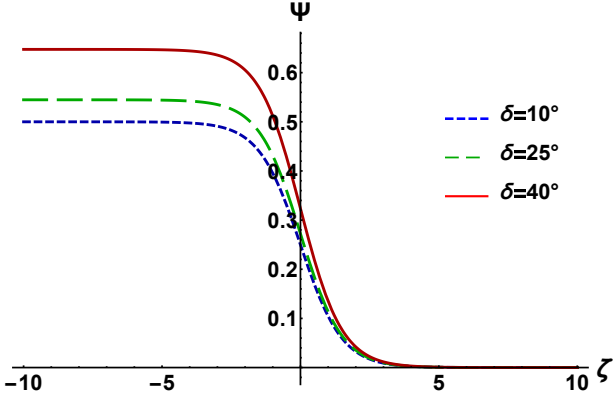}
\caption{The variation of $\Psi$ with $\zeta$ for different values of $\delta$
under consideration of $\mu_e > \mu_{ec}$. Other plasma parameters are $\alpha_1 = 10^{-3}$, $\alpha_2 = 0.03$,
$\eta = 0.3$, $\kappa=2$, $\mu_e = 0.8$,  $U_0 = 0.2$, and $v_p\equiv v_{p+}$.}
\label{2Fig:F2}
\end{figure}
\begin{figure}
\centering
\includegraphics[width=80mm]{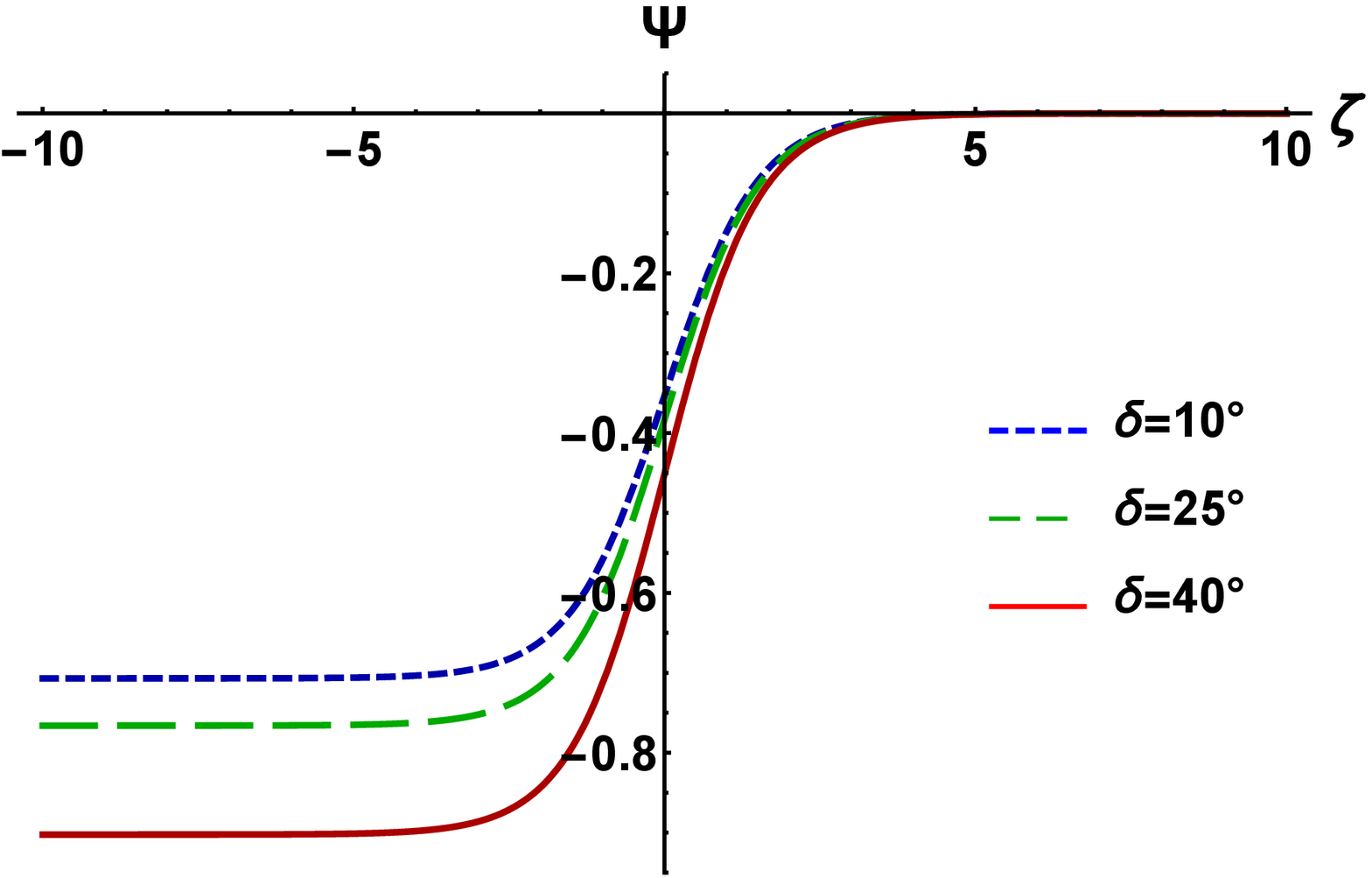}
\caption{The variation of $\Psi$ with $\zeta$ for different values of $\delta$
under consideration of $\mu_e < \mu_{ec}$. Other plasma parameters are $\alpha_1 = 10^{-3}$, $\alpha_2 = 0.03$,
$\eta = 0.3$, $\kappa=2$, $\mu_e = 0.4$, $U_0 = 0.2$, and $v_p\equiv v_{p+}$.}
\label{2Fig:F3}
\end{figure}
\begin{figure}
\centering
\includegraphics[width=80mm]{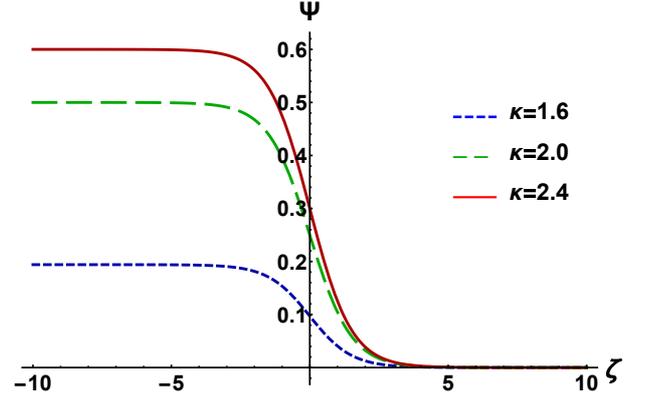}
\caption{The variation of $\Psi$ with $\zeta$ for different values of $\kappa$
under consideration of $\mu_e > \mu_{ec}$.
Other plasma parameters are $\alpha_1 = 10^{-3}$, $\alpha_2 = 0.03$, $\delta=10^\circ$,
$\eta = 0.3$, $\mu_e = 0.8$, $U_0 = 0.2$, and $v_p\equiv v_{p+}$.}
\label{2Fig:F4}
\end{figure}
\section{Numerical analysis}
\label{2sec:Numerical analysis}
The balance between nonlinearity and dissipation leads to generate DIASHWs in a three-component magnetized DPM.
We have numerically analyzed the variation of $A$ with $\mu_e$ in Fig. \ref{2Fig:F1},
and it is obvious from this figure that (a) $A$ can be negative, zero, and positive depending on the values of $\mu_e$;
(b) the value of $\mu_e$ for which $A$ becomes zero is known as critical value of $\mu_e$ (i.e., $\mu_{ec}$),
and the $\mu_{ec}$ for our present analysis is almost $0.55$;
and (c) the parametric regimes for the formation of positive (i.e., $\psi>0$) and negative (i.e.,  $\psi<0$) potential shock structures
can be found corresponding to $A>0$ and $A<0$.

It is clear from Figs. \ref{2Fig:F2} and \ref{2Fig:F3} that (a) with the increase of the oblique angle ($\delta$),
the magnitude of the amplitude of the positive and negative shock structures increases, and this result agrees with the result of Hossen \textit{et al}. \cite{Hossen2017}; (b) the magnitude of the negative potential is always greater than the positive potential for
same plasma parameters. So, the oblique angle enhances the
height of the potential shock structures.
\begin{figure}
\centering
\includegraphics[width=80mm]{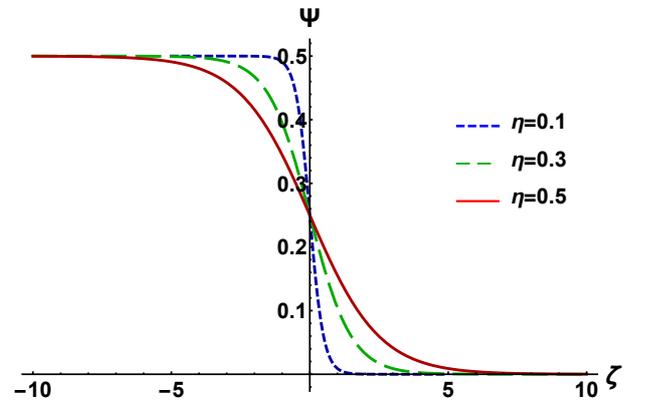}
\caption{The variation of $\Psi$ with $\zeta$  for different values of $\eta$
under consideration of $\mu_e > \mu_{ec}$. Other plasma parameters are $\alpha_1 = 10^{-3}$, $\alpha_2 = 0.03$,
$\delta=10^\circ$, $\kappa=2$, $\mu_e=0.8$, $U_0=0.2$, and $v_p\equiv v_{p+}$.}
\label{2Fig:F5}
\end{figure}
\begin{figure}
\centering
\includegraphics[width=80mm]{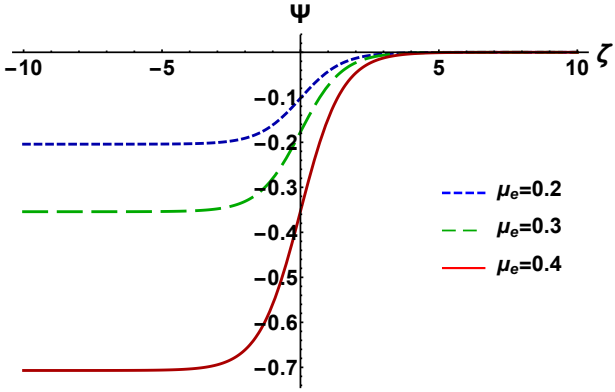}
\caption{The variation of $\Psi$ with $\zeta$ for different values of $\mu_e$
under consideration of $\mu_e < \mu_{ec}$. Other plasma parameters are $\alpha_1 = 10^{-3}$, $\alpha_2 = 0.03$,
$\delta=10^\circ$, $\eta = 0.3$, $\kappa = 2$, $U_0 = 0.2$, and $v_p\equiv v_{p+}$.}
\label{2Fig:F6}
\end{figure}

Figure \ref{2Fig:F4} displays the variation of positive shock profile with spectral index $\kappa$,
and it is evident from the figure that (a) shock height is amplified for higher values of $\kappa$,
and this result agrees with the previous work of Ref. \cite{Shah2012};
(b) positive shock height changes abruptly with the increase of the value of $\kappa$.

It is evident from Fig. \ref{2Fig:F5} that there are some specific
correlations between the dust-ion kinematic viscosity on
the positive (under the consideration $\mu_e>\mu_{ec}$) shock profiles.
It is really interesting that the magnitude of the amplitude of positive shock profiles is not
affected by the variation of the dust-ion kinematic viscosity but the steepness of the shock profile decreases with the increase of
dust-ion kinematic viscosity, and this result is in good agreement with the result of Bansal \textit{et al}. \cite{Bansal2020} and
Borah \textit{et al.} \cite{Borah2016}.

Figure \ref{2Fig:F6} illustrates the effects of the equilibrium density ratio of electrons to ions (via $\mu_e$)
on plasma shock structures. The numerical analysis exhibits amplification in the magnitude of the
amplitude of negative shock profile for higher
values of $\mu_e$ which fully agrees with the result of Kamran \textit{et al.} \cite{Kamran2021}. However, the variation of
the value of $\mu_e$ drastically changes the height of the shock profile.
\section{Conclusion}
\label{2sec:Conclusion}
We have studied DIASHWs in a three-component magnetized DPM by considering kinematic viscosities
of both negative dust and positive ion species, and inertialess super-thermal electrons.
The reductive perturbation method is used to derive the Burgers' equation. The results that have been found from
our investigation can be summarized as follows:
\begin{itemize}
  \item The parametric regimes for the formation of positive (i.e., $\psi>0$) and negative (i.e., $\psi<0$)
  potential shock structures can be found corresponding to $A>0$ and $A<0$.
  \item The magnitude of the amplitude of positive and negative
            shock structures increases with the oblique angle ($\delta$).
  \item The magnitude of the amplitude of positive and negative shock profiles is not affected by the
          variation of the dust-ion kinematic viscosity ($\eta$) but the steepness of the shock profile
           decreases with dust-ion kinematic viscosity ($\eta$).
\end{itemize}
It should be noted here that the gravitational effect is of great importance for DPM but it is beyond
the scope of our present work. In future for
better understanding, someone can investigate the nonlinear
propagation in a three-component DPM by considering the gravitational effect.
The results of our present investigation will be useful
in understanding the nonlinear phenomena both in
astrophysical environments such as interstellar clouds \cite{Spitzer1978,Savage1979,Herbst1995}, circumstellar clouds \cite{Ossenkopf1992,Andrews2005,Dijkstra2005},
interplanetary space \cite{Mackinnon1987,Sandford1989,Liou1996}, Earth’s magnetosphere\cite{Horanyi1987},
Saturn's B ring \cite{Northrop1983} and magnetosphere \cite{Humes1980,Northrop1983a}, Jupiter's magnetosphere \cite{Northrop1983a,Humes1974},
cometary tails \cite{Mazets1987,Gurnett1986} and in laboratory plasmas \cite{Selwyn1989,Boufendi2002}.

\end{document}